\documentclass[12pt,twoside]{article}
\usepackage[mathscr]{eucal}
\usepackage{amsmath,amsfonts,amssymb,amsthm}
\usepackage{hyperref}
\usepackage{times}
\usepackage{srcltx}
\usepackage[normalem]{ulem}

\newcommand{\changed}[1]{{#1}}

\def\sideremark#1{\ifvmode\leavevmode\fi\vadjust{\vbox to0pt{\vss
 \hbox to 0pt{\hskip\hsize\hskip1em
 \vbox{\hsize3cm\tiny\raggedright\pretolerance10000
  \noindent #1\hfill}\hss}\vbox to8pt{\vfil}\vss}}}

\def\be{\begin{equation}}
\def\ee{\end{equation}}
\def\ba{\begin{array}}
\def\ea{\end{array}}

\def\dps{\displaystyle}

\def\sd{S^\dagger}
\def\csd{{\cS}^\dagger}
\def\bsd{\bar S}
\def\cbsd{\bar \cS}

\def\BGST{Barnich:2004cr}
\def\BGadS{Barnich:2006pc}

\def\AGT{Alkalaev:2008gi}
\def\AG{Alkalaev:2009vm}

\advance\textheight\topskip
\renewcommand{\tilde}{\widetilde}
\renewcommand{\hat}{\widehat}
\newtheorem{prop}{Proposition}[section]
\newtheorem{lemma}[prop]{Lemma}

\newcommand{\bref}[1]{\textbf{\ref{#1}}}

\newcommand{\gh}[1]{\mathrm{gh}(#1)}

\newcommand{\smallq}{q}
\newcommand{\smalls}{s}

\newcommand{\dd}{\partial}
\renewcommand{\d}{\partial}

\renewcommand{\geq}{\,{\geqslant}\,}
\renewcommand{\leq}{\,{\leqslant}\,}

\newcommand{\inner}[2]{\langle #1{,}\,#2\rangle}
\newcommand{\binner}[2]{%
  {\langle}\kern-4.15pt{\langle}#1{,}\,#2{\rangle}\kern-4.15pt{\rangle}}
\newcommand{\commut}[2]{[#1{,}\,#2]}

\newcommand{\half}{\mathchoice{%
    \ffrac{1}{2}}{\frac{1}{2}}{\frac{1}{2}}{\frac{1}{2}}}

\newcommand{\ffrac}[2]{\raisebox{.5pt}%
  {\footnotesize$\displaystyle\frac{#1}{#2}$}\kern1pt}

\newcommand{\brst}{\mathsf{\Omega}}

\newcommand{\dl}[1]{\mathchoice{\ffrac{\dd}{\dd #1}}{\frac{\dd}{\dd
      #1}}{\ffrac{\dd}{\dd #1}}{\ffrac{\dd}{\dd #1}}}

\newcommand{\ddl}[2]{\ffrac{\dd #1}{\dd #2}}

\def\const{\mathop\mathrm{const}\nolimits}

\newcommand{\fR}{\mathbb{R}}

\newcommand{\bundle}{\boldsymbol}

\def\cH{\mathcal{H}}

\def\cM{\mathcal{M}}
\def\cN{\mathcal{N}}

\def\cQ{\mathcal{Q}}

\def\cS{\mathcal{S}}

\def\cZ{\mathcal{Z}}

\numberwithin{equation}{section} \makeatletter
\@addtoreset{equation}{section}

\def\weight{w}
\def\uu{u}

\begin{document}

\begin{flushright}
FIAN-TD-2011-08 \\
\end{flushright}

\begin{center}

{\Large\textbf{Unified BRST approach to (partially) massless and
\\ [8pt]
massive AdS fields of arbitrary symmetry type
}}

\vspace{.9cm}

{\large Konstantin Alkalaev and  Maxim Grigoriev}

\vspace{0.5cm}

\textit{I.E. Tamm Department of Theoretical Physics, \\P.N. Lebedev Physical
Institute,\\ Leninsky ave. 53, 119991 Moscow, Russia}

\vspace{0.5cm}

\begin{abstract}

We construct a concise gauge invariant formulation for massless,
  partially massless, and massive bosonic AdS fields of arbitrary
  symmetry type at the level of equations of motion.  Our formulation
  admits two equivalent descriptions: in terms of the ambient space
  and in terms of an appropriate vector bundle, as an explicitly local
  first-order BRST formalism.  The second version is a parent-like
  formulation that can be used to generate various other formulations
  via equivalent reductions.  In particular, we demonstrate a relation
  to the unfolded description of massless and partially massless fields.

\end{abstract}

\end{center}






\section{Introduction}

Arbitrary AdS fields can be divided into three classes according
to particular values of their vacuum energy $E_0$. These are
massive fields, massless fields and partially-massless fields that
carry intermediate number of degrees of freedom. In the simplest
case of totally symmetric fields the above three classes were
described
in~\cite{Fronsdal:1978vb,Vasiliev:1986td,Metsaev:1995re,Deser:1983mm,Deser:2001pe,Zinoviev:2001dt}.
Starting from five dimensions the totally symmetric fields are
only a special case of mixed symmetry ones.

Various approaches to
mixed symmetry fields are known by now. Particularly relevant for
us is the manifestly AdS invariant
formulation~\cite{Metsaev:1997nj} in terms of AdS tensors, which
can be seen as a generalization of the Fronsdal
approach~\cite{Fronsdal:1978vb} to totally symmetric AdS fields.
Another related development has to do with the frame-like
description operating  with  $o(d-1,2)$-valued  $p$-forms as
fundamental fields~\cite{Lopatin:1988hz,Vasiliev:2001wa,Alkalaev:2003qv,Skvortsov:2006at,BIS,Skvortsov:2009zu}.
There are other interesting approaches to mixed-symmetry AdS
fields
~\cite{Metsaev:1999ui,Metsaev:2004ee,Buchbinder:2007ix,Fotopoulos:2008ka,
Reshetnyak:2008sf,Zinoviev:2009gh,Bastianelli:2009eh}. Studying
massive mixed symmetry AdS fields is mainly motivated by the presence
of such fields in the spectrum of strings on AdS~\cite{Tseytlin:2002gz}
(see also~\cite{Bonelli:2003zu,Sagnotti:2003qa} for the string-inspired approach to AdS fields).

In this paper we take a different route and extend our previous
results on unitary massless mixed-symmetry
fields~\cite{Alkalaev:2009vm} to the general case including
non-unitary massless, partially massless, and massive fields.~\footnote{Unitary fields in AdS space are described by
infinite-dimensional $o(d-1,2)$ UIRs with lowest energies saturating the
  \changed{unitarity} bound $E_0 \geq E_0(s_p,d) = s_p - p +d-2$,
where $s_p$ \changed{denotes the $p$-th spin weight}~\cite{Breitenlohner:1982jf}, \cite{Fronsdal:1978vb},\cite{Metsaev:1995re}.
They are called unitary massless and unitary massive fields. Fields with  energies below the boundary
value $E_0(s_p,d)$ are called non-unitary and
include non-unitary massless, non-unitary massive and all partially massless fields. }
Besides the well-known developments in the unfolded formulation of
higher-spin dynamics~\cite{Lopatin:1988hz,Vasiliev:2001wa,Vasiliev:2003ev}
(for a review see~\cite{Bekaert:2005vh}), the approach
of~\cite{Alkalaev:2009vm} and the present paper has its roots in
the so-called parent formulation of~\cite{Barnich:2004cr,
Barnich:2006pc,Grigoriev:2006tt} and the description~\cite{\AGT} of the Minkowski space mixed
symmetry fields.

Our ideology is to keep invariance with respect to AdS algebra
manifest from the very beginning. This is achieved by defining
fields on the ambient space and employing the AdS invariant gauge
equivalence relation. It turns out that from the ambient
perspective it is more natural to use as a parameter weight
$\weight$ determining the radial behavior of a field instead of
the energy $E_0$ which is in fact linearly related to $\weight$.
In this respect, our approach is analogous to the recently
proposed description~\cite{Grigoriev:2011gp} of totally symmetric
fields.\footnote{It is worth mentioning that weight $\weight$ can
be also identified to the weight of tractors involved in the
description in terms of tractor bundles. For further details
see~\cite{Gover} and references therein.}
In terms of weight $\weight$ (partially) massless fields correspond to special
integer values of $\weight$ while massive fields correspond to
generic values of $\weight$. In the later case the gauge
invariance becomes purely algebraic and can be completely
eliminated.

Although the ambient space formulation is very compact and
algebraically transparent its locality is not manifest. The explicitly
local formulation is constructed in the next step by, roughly
speaking, putting the ambient space to the fiber of a bundle over
the genuine AdS space. This step is identical to the one performed
in~\cite{Barnich:2004cr, Barnich:2006pc,\AGT,Bekaert:2009fg} and
from the first quantized point of view amounts to the Fedosov-type
extension~\cite{Fedosov} (see
also~\cite{Grigoriev:2000rn,Batalin:2001je,Barnich:2004cr} for the
generalizations and applications relevant in the present context)
of the starting point system on the ambient space.

The algebraic structure of the proposed formulation is essentially determined by
$o(d-1,2)-sp(2n)$ Howe dual pair~\cite{Howe1} of $AdS_d$
spacetime algebra $o(d-1,2)$ and symplectic algebra $sp(2n)$
realized on the fiber.
In particular, BRST operator of $\Omega = \nabla + Q_p$
is a sum of $o(d-1,2)$ background covariant derivative $\nabla$
built from  $o(d-1,2)$ generators and purely algebraic part $Q_p$
built from  $sp(2n)$ generators while the off-shell constraints
are also expressed through the $sp(2n)$ generators. Besides the
value of $\weight$ the difference between massive, massless, and
partially-massless fields is in the form of the special off-shell
constraint: for massive fields it is not present while for
(partially)-massless fields it is a $t$-th power of the
respective  $sp(2n)$ generator, where $t$ is the ``depth'' of the
partially massless gauge transformation~\cite{Deser:2001pe,Deser:2001us}.

Both the ambient space formulation of AdS
dynamics and its parent-like extension are given at the level of
equations of motion only. The respective Lagrangian formulation is
not constructed yet. While for particular classes of AdS fields
Lagrangian covariant formulation is known in one or another
form~\cite{Brink:2000ag,Alkalaev:2006rw,de Medeiros:2003px,
Skvortsov:2006at,Reshetnyak:2008sf,Zinoviev:2009gh} this is not
the case for general AdS fields. We hope that the algebraic and
geometric structures identified in the present paper will be also
helpful in constructing Lagrangians in the general case. It is
expected that the ambient space BRST Lagrangian analogous to the
flat space one of~\cite{\AGT} (see also~\cite{Burdik:2001hj,Sagnotti:2003qa})
determines Lagrangian for AdS fields through a version of the radial reduction
of~\cite{Biswas:2002nk,Bekaert:2003uc,Hallowell:2005np}.

The paper is structured as follows. In the next section we recall
some basic algebraic facts on Howe dual realizations of $o(d-1,2)$
and $sp(2n)$ algebras. Besides the standard realization of the
dual algebras we need the so-called twisted realization. Then in
Section \bref{sec:ambient} we describe AdS fields as ambient space
tensor fields subjected to the appropriate constraints, equations
of motion, and gauge transformations. We then explicitly describe the
choice of the parameter $\weight$ and the extra constraints that
give one or another irreducible system. In Section~\bref{sec:genBRSTform}
the ambient space formulation is lifted to the manifestly local formulation where
the ambient space is promoted to a fiber of the appropriate
vector bundle over the AdS space. Cohomology of the fiber part
$Q_p$ of the BRST operator is analyzed in
Section~\bref{sec:cohomology} where it is shown nonvanishing in
the minimal and the maximal ghost numbers only. The former  is
identified with the gauge module of the respective unfolded
formulation while the later with the Weyl module. Massive fields
are discussed in Section~\bref{sec:massive}. The summary of the obtained results is given in Section~\bref{sec:concl}.
Appendices contain various technical details needed in the main text.

\section{Algebraic preliminaries}
\subsection{Howe dual realizations}
\label{sec:howe}

A usual way~\cite{Fronsdal:1978vb,Metsaev:1995re,Metsaev:1997nj} to describe
fields on AdS space in such a way that the isometry algebra is realized linearly
is to work with tensors of AdS algebra instead of Lorentz tensors. Moreover, it
is also useful  to identify the space-time itself as a hyperboloid embedded
in the flat ambient space so that the isometries are ambient pseudo-orthogonal
transformations. Following~\cite{Alkalaev:2009vm} we now recall
algebraic tools necessary to handle arbitrary fields on AdS space in a unified way.

Let $X^A$, $A = 0,..., d\,$ be Cartesian coordinates on the $d+1$-dimensional ambient space $\fR^{d-1,2}$.
We use the usual identification of  AdS space as a hyperboloid
\begin{equation}
\label{hyper}
\eta_{AB}X^A X^B + 1 = 0\;,
\qquad \eta_{AB} = \mathrm{diag}{(- + \cdots + -)}\;.
\end{equation}
Infinitesimal  isometries of the hyperboloid form a pseudo-orthogonal algebra $o(d-1,2)$.

Let $A^A_I$, where $A=0, ..., d$ and $I=0, ...., n-1$ be commuting variables transforming as
vectors of $o(d-1,2)$. The space of functions in $A^A_I$ is naturally an $o(d-1,2)-sp(2n)$-bimodule.
More precisely, $o(d-1,2)$ is realized by
\begin{equation}
\label{JAB}
J^{AB} =A^A_I\dl{A_B{}_I}-A^B_I\dl{A_A{}_I}\;.
\end{equation}
The realization of $sp(2n)$ reads
\begin{equation}
\label{SPgenerators}
 T_{IJ}=A_I^A A_{JA}\,,
 \qquad
 T_I{}^J=\frac{1}{2}\,\big\{A^A_I, \dl{A_J^A}\big\}\,,
 \qquad
 T^{IJ}=\dl{A_I^A}\dl{A_{JA}}\,.
\end{equation}
These two algebras form a Howe dual pair $o(d-1,2)-sp(2n)$ \cite{Howe1}.
In particular, they commute in this representation. The diagonal elements $T_I{}^I$ form a
basis in the Cartan subalgebra while $T^{IJ}$ and $T_I{}^J,\; I<J$ are the basis
elements of the appropriately chosen upper-triangular subalgebra. Let us note that $gl(n)$ algebra is realized by the generators $T_I{}^J$ as a subalgebra of $sp(2n)$ while
its $sl(n)$ subalgebra is generated by $T_I{}^J$ with $I\neq J$.

In what follows we also need to pick a distinguished direction in the space of oscillators
$A_I^A$. Without loss of generality we take it along $A_0^A$ so that from now on we consider
variables $A^A_0$ and $A^A_i$, $i=1,..., n-1$ separately.
In particular, we identify $sp(2n-2)\subset sp(2n)$ subalgebra preserving the direction.
We use the following notation for some of $sp(2n-2)$ generators
\begin{equation}
\label{glnot}
N_i{}^j\equiv T_i{}^j=A_i^A\dl{A_j^A}\,\,\,\;\;\; i\neq j\,,
\qquad
N_i=N_i{}^i\equiv T_i{}^i-\frac{d+1}{2}=A_i^A\dl{A_i^A}\,,
\end{equation}
which form $gl(n-1)$ subalgebra, and
\begin{equation}
\label{smallTr}
T_{ij} = A_i^A A_j{}_A\;,
\qquad
T^{ij} = \dl{A_i^A}\dl{A_j{}_A}\;,
\end{equation}
that complete the above set of elements to
$sp(2n-2)$ algebra.

In what follows we use two different realizations of $sp(2n)$
generators involving  $A_0^A$ and/or $\d/\d A_0^A$.

\subsubsection{Realization on ambient space functions}
\label{sec:realization-amb}

In this case we take the space of polynomials in $A^A_i$ with coefficients
in smooth functions on $\fR^{d+1}$ with the origin excluded. If $X^A$ are coordinates on $\fR^{d+1}$
the representation for $A_0$ and $\dl{A_0}$ is given by
\begin{equation}
A_0^A = X^A, \qquad \dl{A_0^A}=\dl{X^A}\,,
\end{equation}
while the remaining variables $A^A_i$ are represented as before.

We keep the previous notation \eqref{glnot}, \eqref{smallTr}
for generators that do not involve $X^A$ and/or $\d/\d X^A$ while those that do are
denoted by
\begin{equation}
\label{sp3}
\ba{l}
\dps
\cS^\dagger_i=A_i^A\dl{X^A}\,,\qquad\; {{\bar \cS}}^i=X^A\dl{A_i^A}\,,
\\
\\
\dps
\cS^i=\dl{A^A_i}\dl{X_A}\,,\qquad \Box_X=\dl{X^A}\dl{X_A}\,.
\ea
\end{equation}

\subsubsection{Twisted realization}
\label{sec:twisted}
Another possibility is to realize the dual algebras on the space of polynomials
in $A_i^A$ with coefficients in formal power series in variables $Y^A$ such that
$A_0$ and $\dl{A_0}$ are realized as
\begin{equation}
\label{r21}
A_0^A =  Y^A+V^A\,, \qquad  \dl{A_0^A}=\dl{Y^A}\,,
\end{equation}
where $V^A$
is some $o(d-1,2)$ vector normalized as
$V^AV_A=-1$. The $sp(2n)$ generators involving $A_0$ are then realized by
(inhomogeneous) formal differential operators on the space of ``functions''
in $A^A_i$ and $Y^A$.  We use the following notation
\begin{gather}
\label{sp1}
\ba{l}
\dps
\sd_i=A_i^A\dl{Y^A}\,,\qquad \bsd^i=(Y^A+V^A)\dl{A_i^A}\,,
\\
\\
\dps
S^i=\dl{A^A_i}\dl{Y_A}\,,\qquad \Box_Y=\dl{Y^A}\dl{Y_A}\,.
\ea
 \end{gather}
This realization of the dual orthogonal and symplectic algebras is refereed to as twisted Howe dual realization.

The twisted realization is the same as in \cite{\AGT} but with
$Y^A$ replaced by $Y^A+V^A$. Shifting by $V^A$ is crucial because
this realization is inequivalent with the usual one
(\textit{i.e.}, the one with $V^A=0$). This happens because the
change of variables $Y^A\to Y^A+V^A$ is ill-defined in the space
of formal power series. In contrast to the usual realization where
highest (lowest) weight conditions of $sp(2n-2)$ determine
finite-dimensional irreducible $o(d-1,2)$-modules, the
inhomogeneous counterpart of these conditions can determine both
finite-dimensional irreducible or infinite-dimensional
$o(d-1,2)$-modules. In particular, it allows one to describe
finite-dimensional gauge modules and infinite-dimensional Weyl
modules\footnote{In this case it reduces to the so-called
twisted-adjoint module of~\cite{Vasiliev:2001wa,Vasiliev:2003ev,Alkalaev:2003qv,BIS,Skvortsov:2009nv}.}
associated to AdS gauge fields at the equal footing. Note that
the above realization for $n=1,2$ has been originally described in~\cite{\BGadS}
and in~\cite{\AG} for general $n$. Analogous representation has been also used
in~\cite{Bekaert:2009fg} to describe conformal fields.

\section{Ambient space  description of AdS gauge fields}
\label{sec:ambient}
\subsection{Constraints and gauge symmetries}

Using realization~\bref{sec:realization-amb} in terms of functions on $\fR^{d+1}/\{0\}$ with values in polynomials in $A^A_i$
unitary massless fields on AdS can be formulated in manifestly $o(d-1,2)$
invariant terms \cite{Fronsdal:1978vb,Metsaev:1995re}.\,\footnote{See ref. \cite{Bekaert:2010hk} for a nice review of ambient space formulation
of AdS tensor calculus.} More precisely, the space of field configurations can be described~\cite{Alkalaev:2009vm}
by imposing a certain parabolic subalgebra of $sp(2n)$ followed by taking a quotient with respect to gauge
transformations generated by $\cS^\dagger_\alpha$ with $\alpha=1,\ldots, p$.

Now we extend this description to the case of not necessarily
unitary and massless fields. It turns out, however, that in
general one is forced to allow for higher powers of certain
$sp(2n)$ generators. Constraints to be imposed on the ambient
space field $\phi = \phi(X,A)$ are grouped as follows

\vspace{-5mm}

\paragraph{General off-shell constraints.} These are tracelessness, Young symmetry and spin
weight conditions
\begin{equation}
\label{mainconstr}
T^{ij}\phi = 0\;,
\qquad
N_i{}^j \phi = 0\;\;\;i<j\;,
\qquad
N_i\phi = s_i\phi\;.
\end{equation}
It follows that spin numbers are ordered as $s_1 \geq s_2 \geq ... \geq
s_{n-1}$. To describe generic mixed-symmetry fields  it is sufficient to choose
parameter $n$ satisfying $n\leq [\frac{d+1}{2}]$. In odd dimensions there are
also self-dual fields singled  out by additional constraints involving
Levi-Civita tensor but we do not consider them here.\footnote{For explicit
treatment  of  $AdS_5$ self-dual fields we refer to~Ref. \cite{Metsaev:2004ee}.}

\vspace{-5mm}

\paragraph{Radial constraint.} The radial dependence is fixed by
\begin{equation}
\label{radial}
h\phi=0\,, \qquad h=N_X-\weight\;,\qquad N_X = X^A\dl{X^A}\;,
\end{equation}
where $\weight$ is a real number which serves as a parameter of the theory. This allows
to uniquely represent a field defined on the hyperboloid in terms of the ambient space
field satisfying \eqref{radial}. More explicitly, taking a new coordinate system $(r, x^m)$
in $\fR^{d+1}$, such that $r = \sqrt{-X^2}$ is a radius and $x^m$ are dilation-invariant coordinates
$N_X x^m = 0$, one finds $\phi=\phi_0(x,A)\, r^\weight$.

\vspace{-5mm}

\paragraph{Equations of motion.}
Conditions involving $X^A$-derivatives along the hyperboloid are to be regarded as equations
of motion rather than constraints. These are given by
\begin{equation}
\label{difcond}
\Box_X \phi=0\,,\qquad \cS^i\phi=0\,.
\end{equation}
The last equation may be regarded as a $\dl{X^A}$-transversality condition.

One then postulates a gauge invariance.

\vspace{-5mm}

\paragraph{Gauge invariance.} Let us fix integer parameter $p\leq n-1$ and let
$\chi^\alpha=\chi^\alpha(X,A)$ for $\alpha=1,\ldots,p$ denote gauge
parameter satisfying the gauge parameter version of the above constraints. These
are~\eqref{mainconstr},
\eqref{radial}, and~\eqref{difcond} where the constraints involving $N_i{}^j$,
$N_i$ and $N_X$ are modified as
\be
\ba{l}
\dps
N_i{}^j\chi^\alpha+\delta_i^\alpha\delta^j_\beta \chi^\beta = 0\;\;\;i<j\;,
\\
\\
N_i\chi^\alpha+\delta_i^\alpha \chi^\alpha -s_i\chi^\alpha=0 \;,
\\
\\
(N_X-w-1)\chi^\alpha=0\,.
\\

\ea
\ee
A gauge equivalence is defined by
\begin{equation}
\label{gauge-equiv}
 \phi\,\sim\, \phi+\cS^\dagger_\alpha \chi^\alpha\,.
\end{equation}
or, equivalently, the gauge transformation reads as $\delta_\chi\phi=\cS^\dagger_\alpha \chi^\alpha$.
One can directly check that this equivalence relation is compatible with the
constraints on the field and the gauge parameter. In fact, the consistency is guaranteed
because $\cS_\alpha^\dagger$ and the remaining constraints are generators
of a subalgebra from $sp(2n)$.

\vspace{-5mm}

\paragraph{Tangent constraints.}
In addition, fields are required not to depend on the transversal to the
hyperboloid components of $A_i$  for such values of $i$
that the gauge invariance is preserved. This is achieved by imposing the following constraints:
%
\begin{equation}
\label{tangent}
 \bar\cS^{\,\hat\alpha\,} \phi = 0\;, \qquad \hat\alpha =  p+1, \; ...\; , n-1\;.
\end{equation}

\vspace{-5mm}

\paragraph{Extra constraint.} Depending on a particular value taken by parameter $\weight$
the above system can be either irreducible or reducible. As we are going to see the former happens
for $\weight$ generic while the later corresponds to special values
\begin{equation}
\label{special-w}
 \weight=s_p-p-t\,.
\end{equation}
Here parameter $t$ takes values $t = 1,2,...,t_{\rm max}$,
where $t_{\rm max}=s_{p}-s_{p+1}$. In this case the extra irreducibility conditions
\begin{equation}
\label{newS}
(\bar\cS^{p\,} )^{t} \phi = 0\;,
\end{equation}
are to be imposed.

It is a matter of a direct computation that for such $\weight$
constraints~\eqref{newS} and \eqref{radial} are compatible with the gauge
invariance~\eqref{gauge-equiv}. Compatibility of~\eqref{newS} with the remaining constraints
can be directly checked using the following commutation relations of
$sl(n) \subset sp(2n)$
\begin{equation}
\label{commrel}
[N_j{}^k, \csd_i] = \delta_i{}^k \csd_j\;,\qquad [\cbsd{}^i, N_j{}^k] = -\delta^i{}_j \cbsd{}^k\;.
\end{equation}
It follows $N_j{}^k$ with $j<k$ decrease a value of index $i$ for $\csd_i$ towards its minimal value $i=1$
and increase it for $\cbsd^i$ towards its maximal value $i=n-1$. The subalgebra
formed by constraints and gauge generators may therefore involve Young symmetrizers $N_i{}^j$ along with both $\csd_\alpha$
and $\cbsd^{\hat \alpha}$ where $\alpha$ necessarily starts with $\alpha =1$ and $\hat \alpha$ necessarily ends up
with $\hat \alpha = n-1$.

Furthermore, setting  $t=1$ and $s\equiv s_1 = s_2 = ... = s_p$ amounts to describing
unitary gauge fields.
To relate the present discussion to~\cite{Alkalaev:2009vm} let us note that constraints
\eqref{tangent} and \eqref{newS}
are equivalent to imposing  $\cbsd^i \phi = 0$, where index $i$ runs all admissible
values, $i=1,...,n-1$. This happens because in this case the respective Young tableaux
have the uppermost rectangular block of the length $s$ and height $p$ so that fields
automatically satisfy $N_\alpha{}^\beta = 0$ for any $\alpha\neq  \beta$ and hence $\cbsd^p\phi=0$ implies
$\cbsd^\alpha\phi=0$.

\subsection{Ghost variables and BRST operator}
The constraints for both field and gauge parameter can be compactly formulated
if one introduces Grassmann odd ghost variables $b_\alpha$ with ghost number
$\gh{b_\alpha}=-1$. In terms of
generating functions $\Psi(X,A|\,b)$ those constraints
from~\eqref{mainconstr}\,-\,\eqref{difcond} that do not involve
$N_i{}^j,N_i,N_X$ stay intact while the remaining ones take the form
\begin{equation}
(N_i{}^j+B_i{}^j) \Psi = 0\;\;\;i<j\;,
\qquad
(N_i+B_i)\chi = s_i\Psi\;,
\qquad
(N_X-B-\weight)\Psi=0\,.
\end{equation}
Here the following notation for ghost contributions have been introduced:
\begin{equation}
B_i{}^j =\delta_i^\alpha\, \delta^j_\beta\, b_\alpha\dl{b_\beta}\;,
\qquad  B_\alpha  = b_\alpha\dl{b_\alpha}\,,
\qquad
B= \sum_{\alpha=1}^p B_\alpha\,.
\end{equation}
It is easy to see that for zeroth ghost degree component  $\Psi^{(0)}=\phi(X,A)$
and  for  degree minus one  component $\Psi^{(-1)}=\chi(X,A|\,b)=\chi^\alpha(X,A)b_\alpha$ the
constraints for fields and gauge parameter are reproduced.
At the same time the gauge transformation
takes the usual form
\begin{equation}
\label{gauge-sym}
\delta \phi=\cQ_p \chi\,,\qquad \cQ_p=\cS^\dagger_\alpha \dl{b_\alpha}\,,                                              \end{equation}
where $\cQ_p$ is a BRST operator, $\gh{\cQ_p}=1$.

\subsection{Interpretation of parameters}

Our theory is determined by several parameters
which are spins   $s_1 \geq s_2\geq  ...\geq  s_{n-1}$, real parameter $\weight$, integer parameter $p$ entering
the formulation through the gauge equivalence~\eqref{gauge-equiv} and constraints~\eqref{tangent}.
In addition, for special values of $\weight$ extra integer parameter $t$, the ``depth'' of gauge transformations
also shows up through the constraint~\eqref{newS}.

To see which representation we are dealing with let $\Phi(X,A)$ represent
an equivalence class of field configurations modulo the gauge equivalence generated
by $\cQ_p$, \textit{i.e.}, $\Phi \sim \Phi+ \cQ_p\chi$ with $\chi=b_\alpha \chi^\alpha$.
We then explicitly evaluate the value of the quadratic Casimir operator
\begin{equation}
 C_2=-\half J_{AB}J^{AB}\,,\qquad J_{AB}=L_{AB}+M_{AB}\,,
\end{equation}
where the orbital and the spin parts are given by
\begin{equation}
L_{AB}=X_A\dl{X^B}-X_B\dl{X^A}\,, \qquad
M_{AB}=\sum_{i=1}^{n-1}\Big( A_{Ai}\dl{A_i^B}-A_{Bi}\dl{A^A_i}\Big)\,.
\end{equation}
Taking into account constraints~\eqref{mainconstr}-\eqref{difcond} a direct calculation yields
\begin{equation}
\label{inter}
 C_2\Phi=\Big(\weight(\weight+d-1)+\sum_{l=1}^{n-1}s_l(s_l-2l+d-1)\Big)\Phi-2\sum_{l=1}^{n-1}\cS^\dagger_l{\bar \cS}^l\Phi\,.
\end{equation}

Let us analyze the last term in~\eqref{inter} in more detail.  Summands $\cS^\dagger_l{\bar \cS}^l$ with $l> p$ vanish because of the constraints~\eqref{tangent}. The remaining ones are identically rewritten  as
\begin{equation}
\label{}
 \sum_{\alpha=1}^p \cS^\dagger_\alpha {\bar \cS}^\alpha \Phi=\cQ_p\chi\,, \qquad \chi =b_\alpha {\bar \cS}^\alpha \Phi\;.
\end{equation}
It is easy to see that $\chi$ satisfies all the necessary
constraints provided $\Phi$ does. Indeed, the only nontrivial
point is to check that $(N_i{}^j+B_i{}^j)\chi=0$ but this follows from
$\commut{N_i{}^j+B_i{}^j}{b_\alpha {\bar \cS}^\alpha}=0$, which in turn
is algebraically analogous to $\dps\commut{N_i{}^j+B_i{}^j}{\csd_\alpha
\dl{b_\alpha}}=0$.
Note that for the above argument
to work the entire set of oscillators $A_i$ is split into two complementary parts:
oscillators $A_\alpha,\,\,$ $\alpha=1,\ldots,p$ involved in the gauge transformations
$\delta\Phi=\cS^\dagger_\alpha \chi^\alpha$ and oscillators $A_{\hat\alpha},\,\,$ $\hat\alpha=p+1,\ldots,n-1$
entering tangent constraints ${\bar \cS}^{\hat \alpha}\Phi=0$. In other words the respective Young tableau
is cut into two complementary parts: the upper part subjects to the gauge equivalence and the lower part subjects
to the tangent constraints.

It follows that the last term in \eqref{inter}  is pure gauge
(cohomologically trivial) and does not  contribute to the value of the Casimir operator  understood as acting
on equivalence classes of field configurations modulo gauge transformations. In a more refined
language what we have just computed is the value of the second Casimir operator in the $\cQ_p$-cohomology
at zeroth ghost degree. This is a well-defined problem because $C_2$ commutes with $\cQ_p$ as well
as with all the constraints and hence acts in the cohomology.

All in all, one  obtains
\begin{equation}
 C_2\Phi=\big(\weight(d-1+\weight)+\sum_{l=1}^{n-1}s_l(s_l-2l+d-1)\big)\Phi\,,
\end{equation}
so that the analysis in terms of gauge equivalence classes gives the same result as
the gauge fixed analysis of~\cite{Metsaev:1995re,Metsaev:1997nj}. Again
following~\cite{Metsaev:1995re,Metsaev:1997nj} we compare the obtained value with the known value of $C_2$
in the representation with energy $E_0$ and the same spin.
This gives the following identification
\begin{equation}
 E_0(E_0-d+1)=\weight(\weight+d-1)\;,
\end{equation}
so that there are two possible energy values
\begin{equation}
 E^1_0=-w\,, \qquad E_0^2=w+d-1\;.
\end{equation}

Let us discuss two cases separately. If $w$ is special, \textit{i.e.,} $w=s_p-p-t$ one gets
$E^1_0=-(s_p-p-t)$ and $E^2_0=s_p-p-t+d-1$. According to Refs. \cite{Metsaev:1995re,Deser:2003gw,Skvortsov:2009zu}
the correct value of the vacuum of mixed symmetry massless or
partially massless fields is given by $E^2_0$.

If $w$ is generic, the gauge symmetry can be shown purely algebraic so that
there are no genuine gauge fields.  After fixing this algebraic gauge symmetry
one arrives at the formulation without gauge symmetry at all. The respective
field $\tilde \Phi(x,A)$ depending on intrinsic AdS coordinates
$x^m$, where $m=0, ..., d-1$, satisfies the following
equations of motion
\begin{equation}
\tilde \Box  \tilde\Phi=\mu^2\tilde\Phi\,,\qquad \mu^2=\weight(\weight+d-1)\;,
\end{equation}
along with further differential and algebraic
constraints originating from respectively $\cS^i\Phi=0$ and constraints~\eqref{mainconstr}.
Here $\tilde\Box$ is an operator representing $-\half L_{AB}L^{AB}$ in terms of chosen
representatives of equivalence classes. Note that the explicit parameterization of $\tilde\Phi$ and the explicit form of
the $\tilde\Box$ and further constraints depend on the gauge choice
and is discussed in more details in Section~\bref{sec:massive}.

\section{Generating BRST formulation}
\label{sec:genBRSTform}

The formulation based on the ambient space is not manifestly local. Indeed, even
if one explicitly solves the radial constraint and represents fields,
constraints, and gauge transformations in terms of intrinsic coordinates on
AdS space the gauge parameter is still subjected to the differential constraints
(besides the purely algebraic ones). A possible way out is to use a BRST
first-quantized technique and to impose the constraints involving $X^A$-derivatives
through the BRST procedure by adding them to the ``minimal'' BRST operator
$\cQ_p$ with their own ghost variables (see the discussion in Section~\bref{sec:parent}).
This extends the spectrum of fields and ensures that the gauge parameter is not subjected
to differential constraints. Another, though equivalent, approach is to enlarge the space of fields in a more
geometrical way by putting the ambient space to a fiber of a vector bundle over
AdS space~\cite{\BGST,\BGadS,Bekaert:2009fg}. Here we follow the respective
considerations in~\cite{Alkalaev:2009vm} and hence skip details.

\subsection{Space of fields and BRST operator}

A well-known and extremely useful way to describe AdS
geometry\footnote{Analogous technique is easily extended to
generic constant curvature and conformal spaces or, more
generally, parabolic geometries. It is essentially a version of
the well-known Cartan description.} is to consider a trivial vector
bundle $\bundle V$ over $d$-dimensional AdS space with the fiber being the
ambient space $\fR^{d-1,2}$ and the structure group $O(d-1,2)$.
Assume in addition that the bundle is equipped with the flat
$o(d-1,2)$-connection $\omega_m^{AB}(x)$ and a fixed section
$V^A(x)$ satisfying $\eta_{AB}V^AV^B=-1$, where $\eta_{AB}$ is the
standard fiberwise pseudoeuclidean metric \eqref{hyper}~\footnote{Section $V^A$
plays the role of the compensator field ( see, \textit{e.g.},
\cite{compensator})}. If in addition, a local frame
$e^A_m(x)=\nabla_m V^A(x)$ has a maximal rank (\textit{i.e.}, $d$) at any point
then $\omega^{AB}_m(x), V^A(x)$ determine negative constant curvature geometry.
Indeed, $g_{kl}=\eta_{AB}\,e_k^Ae_l^B$ gives the AdS metric. Using
a special local frame where $V^A=(0,\ldots,0,1)$ it is easy to
observe that the flatness condition for $\omega^{AB}_m$ reproduces
the negative constant curvature condition for $g_{kl}$.

In addition we introduce space $\cH$ of
polynomials in $A^A_i$ and ghosts $b_\alpha$ with coefficients in
formal power series in variables $Y^A$ and where the $sp(2n)$ and
$o(d-1,2)$ algebras are given in a twisted realization as explained
in section ~\bref{sec:twisted}. For the moment we do not take explicitly
into account the fiber version of the
constraints~\eqref{mainconstr}\,-\,\eqref{difcond}, and \eqref{newS} because now they are purely algebraic and
concentrate first on the relevant geometrical structures.

The vector bundle we are interested in is a vector bundle associated to $\bundle V$ and with the fiber being $\cH$.
In particular, the flat connection $\omega_m^{AB}$ determines a flat covariant derivative
\begin{equation}
\label{nabla}
\nabla=\theta^m\dl{x^m}+\half\theta^m\omega^{AB}_{m}J_{AB} \,,
\end{equation}
where $J_{AB}$ are $so(d-1,2)$ generators~\eqref{JAB} acting on $\cH$ in a twisted realization and we have
assumed the local frame where $V^A=\const$. Here and below we replace basis differentials $dx^m$ on AdS space
with the Grassmann odd ghost variables $\theta^m$, $m=0,...,d-1$, $\gh{\theta^m}=1$ because in the BRST formulation
$\nabla$ appears as a part of BRST operator.

The BRST extended space of states $\bundle\cH$ is given by differential forms of all ranks on AdS
with values in the bundle. In plain terms they are $\cH$-valued fields depending on $x^m,\theta^m$.
Note that from the first-quantized point of view
the space of field configurations is the BRST extended space of quantum states. The component fields
entering $\Psi = \Psi(x,\theta|Y,A,b)$ have
the following structure
\begin{equation}
\label{component}
\Psi_{m_1 ... m_r}{}^{A_1,...\, ,\,A_l\,,  ...  \,;\,}{}^{\alpha_1 ... \alpha_k}(x)\;,
\end{equation}
where $A_l$ are $o(d-1,2)$ vector indices while $\alpha_k$ and $m_r$ are antisymmetric indices
because the respective ghost variables $b_\alpha$ and $\theta^m$ are anticommuting.

On the space of states $\bundle\cH$ we define the following BRST operator
\begin{equation}
\label{fiberQ}
\hat \brst = \nabla+Q_p\;.
\end{equation}
Here  $\nabla$ is the covariant derivative~\eqref{nabla} and $Q_p$ is the algebraic operator
given by
\begin{equation}
\label{Qp}
Q_p=\sd_\alpha\dl{b_\alpha}\,\;,
\end{equation}
where  $\sd_\alpha$ are $sp(2n)$ generators~\eqref{sp1}. Of course $Q_p$ is precisely
the fiber version of the ambient space BRST operators $\cQ_p$ from~\eqref{gauge-sym}.

Because of the ghost degree prescription $\gh{\theta^m}=-\gh{b_\alpha}=1$ BRST operator
$\hat\brst$ has a standard ghost degree $\gh{\hat\brst}=1$. Moreover,
it follows from $\nabla^2=0$, $Q_p^2=0$ and $o(d-1,2)-sp(2n)$-bimodule structure according to which $J_{AB}$
commutes with all the $sp(2n)$ generators that $\hat\brst$ is nilpotent so that it can be consistently
interpreted as a BRST operator.

\subsection{Fiber constraints and equations of motion}
\label{sec:const+eom}

Before discussing equations of motion and gauge symmetries we need to
impose the fiber version of the constraints introduced in the ambient space
description of Section~\bref{sec:ambient}.  More precisely, off-shell
constraints
\begin{equation}
\label{mainconst-fiber}
T^{ij}\Psi = 0, \qquad (N_i{}^j+B_i{}^j) \Psi = 0\;\;\;i<j,\qquad
(N_i+B_i)\Psi = s_i\Psi\,,
\end{equation}
stay the same while those involving $A_0^A$ (\textit{i.e.}, \eqref{difcond} and \eqref{radial}, \eqref{tangent})
take the form
\begin{equation}
\label{diff-const-fiber}
 \Box_Y\Psi=0\,, \qquad S^i\Psi=0\,,
\end{equation}
and
\be
\label{H}
h\Psi=0\,, \quad\qquad h=N_Y-B-\weight\,,
\ee
\be
\label{tangent-fiber}
\bsd^{\,\hat\alpha\,} \Psi = 0\;, \qquad \hat\alpha =  p+1, \; ...\; , n-1\;.
\ee
Here we recall that $N_Y=(Y^A+V^A)\dl{Y^A}$ and $\bsd^i=(Y^A+V^A)\dl{A^A_i}$ to stress the difference with the ambient space realization. For special values $\weight=s_p-p-t$ one in addition imposes the fiber version of~\eqref{newS}:
\begin{equation}
\label{newS-fiber}
(\bsd^{p\,} )^{t} \Psi = 0\;.
\qquad
\end{equation}
Note that the trace constraints in \eqref{mainconst-fiber} and \eqref{diff-const-fiber}
can be collectively written as $T^{IJ}\Psi = 0$.

That operator $Q_p$ acts in the subspace singled out by the above constraints
follows from the constraint algebra which is identical to the one of the
ambient space description. Covariant derivative  $\nabla$ commutes with all the
constraints because of $o(d-1,2) - sp(2n)$ bimodule structure.

According to the general prescription
the physical fields \footnote{Sometimes the term ``physical''  is used to denote
a minimal covariant field content usually obtained by eliminating auxiliary fields and
Stueckelberg variables.} are identified as elements $\Psi^{(0)}$ at ghost number $0$
and  gauge parameters as elements $\Psi^{(-1)}$ at ghost number $-1$ (see, \textit{e.g.},~\cite{\BGST,\AGT}).
Their component form read off from \eqref{component} is given by $k-l=0$ and  $k-l=-1$, respectively. The equations of motion and the gauge transformations read as
\begin{equation}
\label{eom}
\hat\brst \Psi^{(0)} = 0\;,\qquad \delta \Psi^{(0)} = \hat\brst \Psi^{(-1)}\;.
\end{equation}
The component form of these equations was given in~\cite{\AG}. Reducibility
gauge  parameters are described by ghost-number $-n$ elements and the respective
transformations read as $\delta \Psi^{(-n)} = \hat\brst \Psi^{(-n-1)}$. Elements
of positive ghost degree correspond to the equations of motion and their
(higher) reducibility relations.

Let us comment on the use of the BRST approach in the present context. Usually
the BRST operator is assumed to be hermitian with respect to the inner product
in the representations space. In this case the equations of motion of the
associated free field theory can be derived from a local action of the form
$\inner{\Psi^{(0)}}{\Omega\Psi^{(0)}}$. Throughout this paper we do not require
existence of an inner product and the hermiticity of the BRST operator. From the
field theory point of view this corresponds to working at the level of equations
of motion and their gauge symmetries. This approach was described~\cite{\BGST} to
which we refer for further details.

\subsection{Parent formulation}
\label{sec:parent}

Although the constructed formulation is very compact it is important to stress that
the representation space is highly constrained. A description where (almost)
all the constraints are implemented through the BRST procedure
so that the space of fields is (nearly) unconstrained can be useful.
Formulations of this type are known as parent ones and can be used
to generate other formulations through the elimination of generalized auxiliary fields.

Here we briefly discuss a version of the parent formulation generalizing
the one from~\cite{\AG} and  where all the constraints involving derivatives with respect to $Y^A$
are implemented through a BRST procedure. Namely, parent BRST operator reads as
\begin{equation}
\label{min-parent}
 \Omega^{\rm parent}=\nabla+\bar \Omega\,, \qquad \bar\Omega=Q_p+\text{``more''}=\sd_\alpha\dl{b_\alpha}+c_0\Box+c_i S^i-\delta^i_\alpha c_i \dl{b_\alpha}\dl{c_0}\,,
\end{equation}
where new Grassmann odd ghost variables $c_0$ and $c_i$ have been introduced. Note that the remaining constraints~\eqref{mainconst-fiber} and \eqref{H}-\eqref{newS-fiber} or, more precisely, their $\bar\Omega$-invariant extensions are still imposed directly.

Using this form of the theory one can easily obtain the proper
ambient space BRST description where in contrast to the
formulation of Section~\bref{sec:ambient} gauge parameters are not
subjected to differential constraints. Indeed, following~\cite{\AG} one
shows that the parent theory is equivalent to the ambient space
BRST formulation determined by $\bar\Omega$ where all the
constraints are taken in the ambient space realization
of~\bref{sec:realization-amb} instead of the twisted one from section ~\bref{sec:twisted}. The same
applies to off-shell constraints \eqref{H}-\eqref{newS-fiber}.
\footnote{It is worth mentioning that the structure of the resulting BRST
operator which is just $\bar\Omega$ \eqref{min-parent} realized differently
is very similar to the BRST operator used to describe bosonic
strings and HS fields on the Minkowski space (see,
\textit{e.g.},~\cite{Sagnotti:2003qa,\BGST,\AGT}). We expect this
formulation to be useful in constructing the respective Lagrangian
description and analyzing the spectrum of strings on AdS.}

Implementing the remaining constraints through the BRST operator can be easily performed in the
particular case of unitary massless fields, \textit{i.e.}, where $s_1=\ldots=s_p$. In this case
$\cN_\alpha{}^\beta\Psi=0$ for all $\alpha,\beta$
so that $\bar S^p\Psi=0$ imply $\bar S^\alpha \Psi=0$. Consider the following BRST operator
\begin{equation}
\label{brst-tot}
\bar\Omega^{\rm tot}=Q_p+c_{IJ}T^{IJ}+\nu_i\bsd^i+\mu h+\gamma_j{}^iN_i{}^j+\text{ghost terms}\,,
\end{equation}
where we have introduced ghost variables  $c_{IJ}$, $\nu_i$,
and $\gamma_j{}^i\,\,\, i<j$ associated to constraints $T^{IJ},\bsd^i$ and $N_i{}^j \,\,\, i<j$.
It turns out that the parent theory based on $\bar\Omega^{\rm tot}$ is equivalent to the
starting point formulation based on $Q_p$ and constraints~\eqref{mainconst-fiber}-\eqref{newS-fiber}. The proof
is given in Appendix~\bref{app:sln}.

A few comments are in order. Note that $\bar\Omega^{\rm tot}$-invariant extension of the remaining constraints
$(\cN_i-s_i)\Psi=0$ are imposed directly in the representation space. We do not add these constraints to the BRST operator with their own ghosts
because this in general leads to extra cohomology classes. However, imposing them directly is not a real problem
because the entire representation space decomposes into the direct sum of
eigenspaces associated to different values of $s_i$ and the BRST operator preserves the decomposition.
This makes the space subjected to BRST invariant extensions of $(\cN_i-s_i)\Psi=0$ almost as convenient as
a totally unconstrained space.

As far as the general case is concerned the above arguments are
not immediately applicable. This means that using the appropriate
generalization of~\eqref{brst-tot} can, in principle, bring extra
fields and hence spoil the equivalence. Extending~\eqref{brst-tot}
beyond the unitary case remains an open problem.

\section{$Q_p-$cohomology analysis}
\label{sec:cohomology}
For a system whose BRST operator has the structure $\Omega=\nabla+Q$ with $Q$ algebraic
an important information is encoded in the $Q$-cohomology. In the case at hand
the relevant cohomology is the $Q_p$-cohomology in the fiber, \textit{i.e.}, the subspace of
$\cH_{\rm on-shell} \subset \cH$ determined by constraints~\eqref{mainconst-fiber}-\eqref{newS-fiber}.
The $Q_p$-cohomology is graded by ghost number (note that ghosts $\theta^m$ are not the fiber variables
and hence do not contribute to ghost degree in $\cH$).

The $Q_p$-cohomology can be given various interpretations. First of all,
eliminating all the generalized auxiliary fields associated to elements that are
not in the cohomology one reduces the system to the form where fields take values in
$Q_p$-cohomology only. Such a formulation, known as unfolded
formulation,\footnote{Note that the unfolded approach~\cite{Vasiliev:1988xc} was originally
developed from a different perspective.} is in some sense minimal among the formulations where the space-time derivatives enter only through the de Rham differential. Elements of $Q_p$-cohomology at ghost
degree $-k$ give rise to physical fields which are $k$-forms, gauge parameters
which are $k-1$-forms, etc. In particular, in the context of unfolded approach
$Q_p$-cohomology at vanishing ghost degree is known as Weyl module\footnote{Here, the
term ``module'' refers to a space-time symmetry algebra that is
$o(d-1,2)$ in the present case. $Q_p$-cohomology is an $o(d-1,2)$-module
because $Q_p$ commutes with $o(d-1,2)$ algebra or, more generally, with a space-time symmetry
algebra in question.} while those at negative degree as a gauge module. Their
associated fields are $0$ and $k$-forms and can be related to respectively
linearized curvatures and gauge fields.

The analogue of $Q_p$-cohomology can be identified for a general
gauge theory as well. Starting from a Batalin--Vilkovisky formulation of a
given gauge theory in jet space terms (see, \textit{e.g.},~\cite{Barnich:2000zw} and references therein)
and restricting to the stationary surface (by eliminating contractible pairs for Koszul-Tate part
$\delta$ of the BRST differential) one ends up with the
formulation based on gauge part $\gamma$ of the BRST differential.
Finally, eliminating all the contractible variables for $\gamma$
one reduces the system to the form where the reduced $\gamma$ is
at least quadratic. The remaining variables are the generalized tensor fields and connections
of~\cite{Brandt}, where, in particular, the reduced $\gamma$ has been
explicitly computed for various gauge models including Yang-Mills theory and Einstein gravity.
It turns out that in the
case of linear theory these variables can be identified with the
$Q_p$-cohomology. For instance, generalized tensor fields are
associated to basis elements of the Weyl module while generalized connections to those of the
gauge module. In the case of linear theories this relationship
between the unfolded and the BRST approaches was established
in~\cite{\BGST,\BGadS}. The case of general gauge theories was
considered in~\cite{Barnich:2010sw,Grigoriev:2010ic} to which we refer for the exhaustive
discussion.

In this section we explicitly compute $Q_p$-cohomology in the subspace $\cH_{\rm on-shell}\subset \cH$ determined by
\eqref{mainconst-fiber}\,-\,\eqref{newS-fiber}. Namely, we show that
\begin{equation}
H^k(Q_p,\cH_{\rm on-shell})=
\left\{
\ba{l}
\dps
\text{Weyl module}\;,\quad\;\,  k=0\;,
\\
\dps
\qquad
0
\;, \qquad\qquad   k\neq 0,-p\;,
\\
\text{Gauge module}\;,  \quad k = -p\;.
\ea
\right.
\end{equation}
where $H^k(\cdot)$ denotes cohomology at ghost degree $k$.
The cohomology for intermediate ghost numbers is empty while
for ghost number $k=0$ it is non-vanishing
and can be identified with infinite-dimensional Weyl module.
Cohomology at ghost degree $-p$ is nonvanishing for special values~\eqref{special-w} of parameter $\weight$
only. In this case it describes the finite-dimensional gauge module introduced
within the unfolded formulation~\cite{Vasiliev:2001wa,Alkalaev:2003qv,Alkalaev:2003hc,BIS,Skvortsov:2009zu,Alkalaev:2009vm,Skvortsov:2009nv}.

The unfolded equations are determined by the BRST operator induced by $\hat\Omega$ from~\eqref{fiberQ}
in the cohomology of its second term $Q_p$. This can be computed using the standard homological technique
as explained in~\cite{\BGST,\BGadS,\AGT}. We do not discuss the unfolded form of the equations of motion
and gauge symmetries in further details and refer instead to~\cite{BIS,Skvortsov:2009zu,Ponomarev:2010st}.

The rest of this Section is devoted to the analysis of $Q_p$-cohomology in the case where parameter $\weight$ take
special values $\weight=s_p-p-t$ with $t=1,2, ... ,t_{\rm max}$.
The case of generic $\weight$ corresponds to massive field and is analyzed in Section~\bref{sec:massive}.

\subsection{Minimal ghost degree cohomology: gauge module}
\label{sec:gauge module}
The coboundary condition is trivial at minimal ghost degree
and therefore representatives of $Q_p$-cohomology at ghost degree
$-p$  are defined by the following constraints
\begin{equation}
\label{gaugeconstr}
\ba{c}
\dps
\sd_\alpha\Psi = 0\;,
\qquad (\bsd^{\,p})^t \Psi = 0\;,
\qquad
\bsd^{\hat\alpha}\Psi=0\;,
\qquad (N_{Y}+t-s_p)\Psi = 0\;,
\\
\\
\dps
\alpha =  1, \; ...\; , p\;,
\qquad \quad
\hat\alpha =  p+1, \; ...\; , n-1\;,
\ea
\end{equation}
along with constraints \eqref{mainconst-fiber}. It is useful to describe solutions to
\eqref{gaugeconstr} using the parameterization in terms of
$Y^\prime{}^A= Y^A+V^A$. This change of variables is legitimate because
the first condition in \eqref{gaugeconstr} implies that for a homogeneous component
the degree in $Y^A$ cannot exceed that in $A_i^A$ and hence $\Psi$ is a finite order
polynomial in $Y^A$.

Taking into account the total ghost degree a physical field associated to $Q_p$-cohomology
at ghost degree $-p$ is a differential $p$-form. This is because there should be exactly $p$ variables $\theta^m$
to gain a zeroth total ghost degree.
This $p$-form on AdS space takes values in finite-dimensional irreducible $o(d-1,2)$-module
described by the Young diagram with the following lengths of rows
\begin{equation}
\ba{l}
s_1 -1\geq \ldots \geq s_{p-1}-1
\geq s_p-1 \geq s_{p}-t
\geq s_{p+1}\geq \ldots \geq s_{n-1}\;.
\ea
\end{equation}
Note the row of length $s_p-1$ in the middle of a diagram with
a subsequent row of a length $s_p-t$.
According to \cite{Skvortsov:2009zu} such fields describe spin $(s_1,..., s_{n-1})$ partially massless AdS fields
with the gauge symmetry associated to $p$-th row and having depth $t$.

\subsection{Vanishing of intermediate ghost number cohomology}
\label{sec:vanishingcohomol} The proof that $Q_p$-cohomology
vanishes at intermediate ghost numbers is based on the following observation: a
representative of $Q_p$ cohomology class of intermediate ghost degree can always
be assumed polynomial. This can be shown by using cohomological arguments
starting from the parent BRST operator (see Section~\bref{sec:parent}) implementing all the constraints
$\sd_\alpha,N_i{}^j,h,\bsd^{\hat\alpha},(\bsd^p)^t$ with their own ghosts so
that only the weight conditions are imposed directly. This reformulates the
cohomological problem in the nearly unconstrained space. Note that one needs to
keep in mind that in general this can bring extra cohomology classes. However,
this does not affect the argument.

 Using a suitable degree one then reduces the
problem to the cohomology of the term implementing the constraints
$N_\alpha{}^\beta,\sd_\alpha$ associated to the upper-half of the
respective Young tableaux. The cohomology of the respective terms
are known~\cite{\AGT} and relevant representatives can be chosen
polynomial. One then shows that completion of such elements to
representatives of the total BRST operator can be also taken
polynomial so that representatives can indeed be assumed
polynomial. Another way to see that representatives can be assumed polynomial
is to perform a direct analysis of the respective cocycle condition using the
algebraic technique developed in~\cite{\AG}.

In the space of polynomials it is then legitimate to use a new
variable $Y^\prime{}^A  = Y^A+V^A$ and hence to reformulate the
problem as that of standard finite-dimensional $sl(n)$-modules. In
this way one finds that for a polynomial element of ghost degree
$-k$ constraints \eqref{H} and \eqref{newS-fiber} are in general
inconsistent. Indeed,  we obtain weight conditions
\begin{equation}
\label{weigths}
N_{Y} \Psi_k = (s_p+k-t-p)\Psi_k\;,
\qquad
(N_p+B_p) \Psi_k = s_p\Psi_k\;,
\end{equation}
along with
\begin{equation}
\big(\bsd^{p}\big)^t \Psi_k = 0\;.
\end{equation}
The last condition tells us that $\# {Y^\prime{}^A}  \geq  \# A^A_p - t+1$
 and this contradicts \eqref{weigths} except for $k = p$ because $\# A^A_p$ is either $s_p$ or $s_p-1$ depending on whether
ghost $b_p$ is present or not. In this way we arrive at
\begin{prop}
The $Q_p$-cohomology evaluated in the subspace singled out by constraints
\eqref{H} and \eqref{newS-fiber} is empty in the ghost numbers $\;0<-k< p\,$.
\end{prop}

In the space of formal power series there is also nontrivial
cohomology at ghost degree $0$. This is the Weyl module which we describe in the next Section.

\subsection{Cohomology at vanishing ghost degree: Weyl module}
\label{sec:weylmodule}

The structure of the Weyl module for unitary massless AdS gauge
fields was described in~\cite{BIS,Alkalaev:2009vm}
(see also~\cite{Alkalaev:2003qv} for early analysis and~\cite{Lopatin:1988hz,Vasiliev:2001wa}
for the case of totally symmetric fields) and then for
the general case involving partially massless and non-unitary massless
fields in~\cite{Skvortsov:2009nv}. Just like in the case of unitary fields
the generating BRST formulation gives an independent definition of the
Weyl module as $Q_p$-cohomology at zeroth ghost degree.  In this
way the module structure is implemented in the construction from the
very beginning because $Q_p$ is $o(d-1,2)$-invariant. Moreover,
because the cocycle condition is trivial in this case the Weyl
module is just a quotient of the $o(d-1,2)$-module $\cH_{\rm
on-shell}$ modulo the $o(d-1,2)$-invariant subspace.
Recall that $o(d-1,2)$ algebra is realized in the twisted from, see section~\bref{sec:twisted}.

\subsubsection{Lorentz covariant basis}
\label{sec:lorentz}

We choose a local frame where $V^A = \delta^A_{d}$.
Set $Y^a = y^a$ and  $Y^{d} =z$. Analogously, $A^a_i = a_i^a$ and
$A^d_i = \uu_i$. In what follows, we always assume that all elements $\Psi = \Psi(Y,A)$
are totally traceless, $T^{IJ}\Psi = 0$. Lemma \bref{prop:isom} formulated
in Appendix \bref{sec:AppendixA} shows how constraints
${\bsd}{}^i$ and $h$ fix one or another type of dependence on
$(d+1)$-th variables $z$ and $\uu_i$. In particular, using lemma \bref{prop:isom}
one can represent elements satisfying~\eqref{H}-\eqref{newS-fiber} as series in $\uu_\alpha$ variables
\be
\label{decW}
\psi  =\sum_{k\geq 0} \uu_{ \alpha_1} ... \uu_{ \alpha_k}\,\psi_k^{\alpha_1 ... \alpha_k}(a,y|b)\;.
\ee
The above series terminates at some finite order defined by spins  $s_{\alpha}$ and depth
$t$. It follows that elements $\psi$ do not depend on $\uu_{\hat \alpha}$ and
$(\uu_p)^{t+m}$ for $m\geq 0$.
In addition, homogeneity in $\uu_{\alpha}$ gives a useful degree called level.

Both BRST operator $Q_p$ and the
constraints~\eqref{mainconst-fiber}, \eqref{diff-const-fiber} can be rewritten in terms of
parameterization~\eqref{decW}. In so doing the trace constraints remain unchanged while
the weight and Young symmetry conditions take the form
\be
\label{reduwheight}
\ba{c}
\dps
\big(n_\alpha+ \uu_\alpha\dl{\uu_\alpha}+B_\alpha  -s_\alpha\big) \psi = 0\;,
\qquad
\big(n_{\hat \alpha}-s_{\hat\alpha}\big)\psi = 0\;,
\ea
\ee
\be
\label{reduYoung}
\ba{c}
\dps
\big(n_{\alpha}{}^{\beta} +\uu_{\alpha}\dl{\uu_{\beta}}+B_\alpha{}^\beta\big) \psi = 0\,,
\qquad
n_{\hat \alpha}{}^{\hat\beta}\psi =0\,,
\qquad
\big(n_{\alpha}{}^{\hat\beta}-\uu_\alpha \bar{s}^{\,\hat \beta} \big)\psi =0\,,
\ea
\ee
where $\bar{s}^{\,\hat \beta}= y^a\dl{a_{\hat \beta}^a}$, and $\alpha<\beta$ and $\hat\alpha<\hat\beta$.
Using then constraint \eqref{H} rewritten in Lorentz terms as
$\dps \big((z+1)\dl{z}+y^a\dl{y^a} -B+p+t-s_p\big)\phi = 0$ allows one  to cast
BRST operator into the following form
\begin{equation}
\label{lorq}
\tilde Q_p =  q_p  - \hat h \, \uu_{\alpha}\dl{b_{\alpha}} \;,
\end{equation}
where $\tilde Q_p$ is operator \eqref{Qp} rewritten in terms of parameterization~\eqref{decW}
and
\begin{equation}
\smallq_p=\smalls^\dagger_\alpha\dl{b_\alpha}\equiv a^a_\alpha\dl{y^a}\dl{b_\alpha}\,, \qquad \hat h = n_y-B+p+t-s_p\,.
\end{equation}
Recall that $(\uu_p)^{t}$ is zero in our subspace and therefore the respective contribution in~\eqref{lorq}
can also be vanishing. In particular, for $t=1$ the term in $Q_p$
proportional to $\uu_p$  vanishes. Note also that for unitary massless fields
all $\uu_\alpha = 0$ as a consequence of $s_1=\ldots=s_p$ so that the reduced operator is simply $q_p$ \cite{Alkalaev:2009vm}.

\subsubsection{Weyl module}
\label{sec:weyl}

First of all we recall that a Poincar\'e Weyl (PW) module of spin
$l_1\geq l_2\ldots \geq l_{n-1}$ \cite{Lopatin:1988hz,Skvortsov:2008vs} can be
defined~\cite{\AGT} as a subspace of $sl(n)$ HW vectors in the
space of polynomials in $y^a$ and $a_i^a$ variables satisfying the
respective weight conditions. One can view a PW module of spin
$l_1\geq l_2 \ldots \geq l_{n-1}$ as a subspace singled out by
$n_i{}^j\tilde\psi=0$, remaining HW conditions $\bar s_i \tilde\psi=0$, weight conditions
$(n_i-l_i)\tilde\psi=0$, and vanishing ghost degree condition $\gh{\tilde\psi}=0$.
Given AdS spin $s_1\geq \ldots\geq s_{n-1}$ a PW module is called admissible
associated if its weights $l_i$  satisfy $l_i = s_i-\nu_i$ where
$\nu_i = 0$, $ i\leq p$ and $ \nu_i \geq  0 $,  $ i>p $ and $\nu_{p+1}+\ldots + \nu_{n-1} \leq s_{p+1}$.


For unitary fields the AdS Weyl module is isomorphic to a direct sum of admissible associated PW
modules. The following Proposition  is a slight generalization of this result.
It turns out that
$H^0(q_p)$ calculated for spin weights $(m_1, ..., m_{n-1})$ and
denoted by $\cM_{0,p,m}$ can be decomposed into a direct sum of
some PW modules.
\begin{prop}
\label{qp}
The zero-ghost-number cohomology $\cM_{0,p,m}$ of
BRST operator $q_p$ evaluated in the subspace \eqref{reduwheight}, \eqref{reduYoung}
is isomorphic to a direct sum of admissible associated
PW modules.
\end{prop}

More detailed discussion of the above  proposition is relegated to Appendix \bref{proof42}.
Spin weights $\{m\}$ of admissible PW modules  are defined by original spins and parameters
$p$ and $t$ through weight constraints \eqref{reduwheight}. Denoting
$H^0(q_p)$ on the $k$-th level (see~\eqref{decW}) as  $\cM^{(k)}_{0,m,p}$ and its spin weights as $\{m\}_k$ we
find that spins are given by
$m_{\hat \alpha} = s_{\hat \alpha}$ for
$\hat \alpha = p+1, ..., n-1$, and $m_\alpha = s_\alpha - k_\alpha$
for $\alpha =1,...,p $ such that $k_1+...+ k_p = k$.

Computation of $Q_p$-cohomology reduces then to inspecting how the second term
in \eqref{lorq} acts in $H^0(q_p)$. One can show that using this term any
level-$k$ element $\psi_k \in \cM_{0,l,p}$ whose degree in $y^a$ is smaller than  $s_1$ can be set to zero. Denoting  the subspace of all such elements from  $\cM^{(k)}_{0,m,p}$ as ${\cZ}^{\,(k)}_{0,m,p}$
we arrive at the component description of AdS Weyl cohomology (see Appendix~\bref{proof43} for more details).
\begin{prop}
\label{prop:WEYL}
AdS Weyl module $\cM_0$ of a given spin is isomorphic to a direct sum of
quotient spaces
\begin{equation}
\cM_0  = \bigoplus_{k\geq 0}\,\bigoplus_{\{m\}_k} \cM^{(k)}_{0,m,p}\;\big/{\cZ}^{\,(k)}_{0,m,p}\;\;,
\end{equation}
where $\{m\}_k$ denotes a set of admissible spin weights on the $k$-th level.
\end{prop}
For unitary fields operator $\hat h$ does not
contribute and $Q_p = q_p$. As a result ${\cZ}^{\,(k)}_{0,m,p} = 0$  and AdS Weyl cohomology is a direct sum
of admissible PW modules. In other words, we reproduce here the Brink-Metsaev-Vasiliev conjecture
put forward in \cite{Brink:2000ag} and proved in \cite{BIS,Alkalaev:2009vm,Skvortsov:2009nv}.
For non-unitary fields AdS Weyl module is not a direct sum of admissible PW modules. As an illustration
in Appendix~\bref{sec:example} we perform the analysis in the particular case of partially massless
totally symmetric fields, \textit{i.e.} when
$n=2\,$, $\,p=1\,$, $\,t\geq 1$.

\section{Massive fields}
\label{sec:massive}

We now consider the case of generic values of $\weight$. A crucial observation is that in this case the gauge
symmetry determined by $Q_p$ is purely algebraic. It implies that the theory is equivalent to the one without
gauge freedom through the elimination of generalized auxiliary fields. The approach taken in this Section is
an extension of that from~\cite{Grigoriev:2011gp} to the case of mixed symmetry fields.
Mention the related considerations in~\cite{BIS}, where the algebraic nature of the gauge invariance
in the massive case was observed within the unfolded framework.

 The essential step
is the following
\begin{prop}
For $\weight$ generic the $Q_p$-cohomology in the space of elements $\Psi(Y,A,b_\alpha)$ satisfying \eqref{mainconst-fiber}-\eqref{tangent-fiber} can be identified with $b_\alpha$-independent elements satisfying in addition $\bsd^\alpha\Psi=0$ so that the entire set of constraints reads as
\begin{equation}
\label{massive-weyl}
 T^{IJ}\Psi=0\,, \quad \bsd^i\Psi=0\,, \quad N_i{}^j\Psi=0 \,\,\,\, i<j\,,\quad h\Psi=0\,,  \quad  (N_i-s_i)\Psi=0\,.
\end{equation}

\end{prop}
It follows from the Proposition that after reducing to $Q_p$-cohomology there are no elements of
negative ghost degree left and hence no gauge fields.

\begin{proof}
Let us consider first $Q_p$-cohomology in the space of elements $\Psi(Y,A,b_\alpha)$ satisfying $h\Psi=0$ only. Using Lemma~\bref{prop:isom} in the sector of $z$ variables only one finds that this space is isomorphic to the subspace of $z$-independent elements and the isomorphism map amounts to simply putting $z$ to zero.
Its inverse is constructed recursively order by order in $z$ (see~\cite{\BGadS,\AG} for more details). In terms of
$z$-independent subspace $Q_p$ is represented by $\tilde Q_p$ given by
\begin{equation}
 \tilde Q_p=q_p+\hat h \uu_\alpha \dl{b_\alpha}\,,
\end{equation}
where as before $\smallq_p=a^a_\alpha\dl{y^a}\dl{b_\alpha}$ and $\hat h= n_y-B-\weight$. In contrast
to the case of special $\weight$ the coefficient in front of the second term never vanishes if $w$ is generic.
Using a suitable degree one reduces the problem to the cohomology of the second term which, in turn,
is isomorphic to $\uu_\alpha, b_\alpha$-independent elements. In fact the reduced BRST operator vanishes
in this case because there are no more ghost variables left so that
$\tilde Q_p$ cohomology can be identified with $\uu_\alpha, b_\alpha$-independent elements.
Restoring $z$-dependence the $Q_p$-cohomology can be identified as the subspace
$\dl{\uu_\alpha}\Psi=0$, $h\Psi=0$, and $\dl{b_\alpha}\Psi=0$.
Note that any $Q_p$-cocycle vanishing at $\uu_\alpha=b_\alpha=0$ is trivial.

The following identification of the above cohomology as a subspace is more useful
\begin{equation}
\label{identification}
 \bsd^\alpha \Psi=0\,, \qquad h\Psi=0\,, \qquad \dl{b_\alpha}\Psi=0\,.
\end{equation}
The two spaces are clearly isomorphic as can be seen by using a version of Lemma \bref{prop:isom}
in the sector of $\uu_\alpha$ variables. Moreover, if $\psi$ and $\psi^\prime$ satisfy
respectively $\dl{\uu_\alpha}\psi=h\psi=0$ and $\bsd^\alpha \psi=h\psi=0$ along with
$\psi_{\uu_\alpha=0}=\psi_{\uu_\alpha=0}^\prime$ then $\psi-\psi^\prime$ is a coboundary.
Indeed, the difference $(\psi-\psi^\prime)$ vanishes at ${\uu_\alpha=0}$ and hence
is trivial in $Q_p$-cohomology.

In order to take other constraints into account one starts with the parent BRST operator
implementing all the constraints
\begin{equation}
 \sd_\alpha,\quad \bsd^{\hat\alpha},\quad N_i{}^j\,\,\;\;i<j\,,
\end{equation}
with their own ghost variables and acting in the subspace of elements satisfying $h^\prime\psi=(N^\prime_i+B_i-s_i)\psi=0$,
where $h^\prime,N^\prime_i$ are operators $h,N_i$ modified by necessary contributions from the extra ghost variables.
Note that in the representation space the ghost variables associated to extra constraints are represented by coordinate
ghosts carrying positive ghost degree in contrast to momenta ghosts $b_\alpha$ representing  ghosts associated to gauge
generators. Although all representations of fermionic ghosts are equivalent this is a standard choice if the ghosts degree
in the representation space is normalized such that $\gh{1}=0$ and physical fields appear at zeroth ghost degree.
As we have already seen the parent reformulation can in general bring extra cohomology classes but they do not affect the
argument.

Using a suitable degree such that $Q_p$ is the lowest degree term of the total BRST operator
one reduces the cohomology problem to the $Q_p$-cohomology identified as a subspace~\eqref{identification}
and then considers the reduced cocycle condition for an element $\Psi$ of vanishing ghost degree. Taking
into account that $\Psi$ is necessarily ghost-independent (because $b_\alpha$ are eliminated there are no variables
of negative ghost degree left) one indeed finds all the constraints~\eqref{massive-weyl} besides the trace
constraints $T^{IJ}\Psi=0$. That the same analysis remains true in the totally traceless subspace
can be seen using the cohomological arguments from~\cite{\BGadS,\AG}.
\end{proof}

The statement can be rephrased by saying that for generic values
of $\weight$ the gauge invariance determined by $Q_p$ is purely
algebraic and that $\bsd^\alpha \Psi=0$ is a proper gauge
condition completely removing the gauge freedom. Indeed, if
$\chi^\alpha$ is a gauge parameter satisfying
\eqref{mainconst-fiber}-\eqref{tangent-fiber}  then
$\bsd^\beta \sd_\alpha \chi^\alpha=0$ implies $\chi^\alpha=0$. In
this gauge the equations of motion are simply
constraints~\eqref{massive-weyl} along with the $\nabla\Psi=0$
where now $\Psi$ is a zero form on AdS space. In particular,
constraints~\eqref{massive-weyl} give an $o(d-1,2)$ covariant
definition of Weyl module in the massive case.\footnote{In the
case of totally symmetric fields the respective Weyl module and
the unfolded formulation were originally studied
in~\cite{Ponomarev:2010st} within a different framework.} Let us
note that the above statement do not directly apply if $\weight$
is integer but not special (more precisely, such that $\hat h$ is
not invertible). In this case there are  still no genuine gauge
fields but the structure of the Weyl module can be different
(see the respective discussion in~\cite{Grigoriev:2011gp}).

We now turn back to the formulation in terms of the ambient space.
Consider the system determined by~\eqref{massive-weyl} where the
constraints are realized on ambient space functions $\Psi(X,A)$
instead of the fiber ones $\Psi(Y,A)$. Using the arguments
from~\cite{\AG} one can show that the system is
equivalent\footnote{More precisely, one needs the arguments given
in Section~3.3 of~\cite{\AG}  restricted to the case where no
gauge freedom is present. The idea is to reformulate the ambient
space theory in the ambient parent form and then pull-back the
covariant derivative to the hyperboloid.} to the above system on
the AdS space. In particular, this shows that
$\bar\cS^\alpha\Psi=0$ are again proper gauge conditions.

To obtain the explicit form of equations of motion in terms of
tensor fields on the hyperboloid one employs the standard
isomorphism between ambient space tensor fields satisfying
$(N_X-w)\Psi=0$ along with $\bar\cS^i\Psi=0$ and
respective tensor fields on the hyperboloid (see,
\textit{e.g.},~\cite{Metsaev:1997nj}). More explicitly such an ambient space
field $\Psi= \Psi(X,A)$ gives rise to AdS tensor field $\psi(x,a_i)$
according to $\psi(x,a_i)=\Psi(X^A(x),\ddl{X^A}{x^m} a_i^m)$, where
$X^A(x)$ describe the embedding in terms of local coordinates on
the hyperboloid.

Under the isomorphism the ambient space operator
$\dl{X^A}-\frac{X_AX^B}{X^2}\dl{X^B}$ is mapped to the Levi-Civita
covariant derivatives on AdS tensor fields (see, \textit{e.g.}, \cite{Metsaev:1997nj,Bekaert:2010hk}).
Using the isomorphism the ambient space constraints $\bar\cS^i\Psi=0$,
$\Box_X\Psi=0$, $\dps(N_X-\weight)\Psi=0$ indeed give rise
to usual massive equations of motion\footnote{Here we make use of
formulas from~\cite{Metsaev:1997nj} relating the ambient and the
intrinsic Laplacians.}
\begin{equation}
 (g^{kl}\nabla_k\nabla_l+\sum_{i=1}^{n-1}s_i)\psi=\weight(\weight+d-1)\psi\,,
 \qquad \nabla^m \dl{a^m_i}\psi=0\,,
\end{equation}
where $g^{kl}$ is the inverse to the AdS metric
$g_{kl}=\eta_{AB}\ddl{X^A}{x^k}\ddl{X^B}{x^l}$. At the same time
the algebraic constraints $T^{ij}\Psi= 0$, $(N_i-s_i)\Psi=0$,
$N_i{}^j\Psi=0 \,\,\,\, i<j$ remain unchanged except one needs to
rewrite them in terms of $\psi(a_i)$ and the AdS metric.
Note that constraints $h\Psi=\bar\cS^i\Psi=0$ are needed for the
isomorphism and do not produce any conditions on AdS tensor
fields.

\section{Conclusions}
\label{sec:concl}

In this paper we have constructed the unified BRST formulation for arbitrary bosonic fields in $AdS_d$
spacetime. The space of field configurations is identified as a subspace in the ambient configuration space
which is naturally an $o(d-1,2)-sp(2n)$ bimodule.
A set of particular constraints needed to describe a given AdS field
depends on its spin weights, the vacuum energy, and the depth of its (partially massless) gauge invariance.
The ambient space formulation for massless fields successfully reproduces the results of Metsaev
\cite{Metsaev:1995re}, while for partially massless and massive fields the proposed set of fields, their
gauge symmetries, and equations of motion are new.

In addition to ambient space description an explicitly local generating BRST formulation is constructed
by, roughly speaking,  treating the ambient space as a fiber of a vector bundle over the AdS space-time.
In this case the $o(d-1,2)-sp(2n)$ bimodule structure is realized on the fiber in a twisted way which is
in contrast to the standard realization employed in the ambient space description. The twisted realization
is essential for the entire construction and can be regarded as a twisted version of Howe duality.

It is important to stress the role played by BRST operator $Q_p$ associated to the particular $sp(2n)$ basis elements.
It encodes a gauge symmetry of the theory, both differential and algebraic in the ambient space formulation, and
pure algebraic in the generating BRST formulation.  In the later case we show that non-empty $Q_p$-cohomology
is identified with the generalized curvatures (Weyl module) and the
generalized connections (gauge module) of the unfolded formulation \cite{Vasiliev:2001wa,Alkalaev:2003qv,Skvortsov:2006at,BIS,Skvortsov:2009zu} reproducing
the set of unfolded fields. The full system of unfolded equations can be explicitly determined
by the BRST differential reduced to $Q_p$-cohomology.

\section*{Acknowledgments}
\label{sec:acknowledgements}

\addcontentsline{toc}{section}{Acknowledgments}

We are grateful to X. Bekaert, E.~Feigin,  A.~Semikhatov, I.~Tipunin, M.~Vasiliev,
and especially to R.~Metsaev, E.~Skvortsov, and A.~Waldron. The work of  KA is supported in part by RFBR grant
11-01-00830 and the Alexander von Humboldt
Foundation grant PHYS0167. The work of MG is supported by the RFBR grant
10-01-00408 and the RFBR-CNRS grant 09-01-93105.

\appendix

\section{BRST Cohomology associated with $sl(n)$-modules
and the equivalence proof.}
\label{app:sln}
We are interested in the cohomology of the BRST operator of the upper-triangular subalgebra of $sl(n-1)$
formed by $N_i{}^j$, $i<j$ \eqref{glnot} with coefficients in a given finite-dimensional representation.
Introducing ghost variables $\gamma_j^i, \, i<j$ the BRST operator reads as
\begin{equation}
\label{slBRST}
\Omega =\gamma_j^i N_i{}^j-\gamma^i_l \gamma^l_j \dl{\gamma^i_j}\,.
\end{equation}
It is defined on the tensor product of an $sl(n-1)$-module with a Grassmann algebra
generated by ghosts $\gamma_j^i$. Restricting operator \eqref{slBRST} to a
subspace of elements  with definite $sl(n-1)$ weights
we introduce the following  BRST extension of elements \eqref{glnot}:
\begin{equation}
\hat N_i=N_i-\gamma^i_k\dl{\gamma^i_k} + \gamma_i^k\dl{\gamma_i^k}\;.
\end{equation}
It is then easy to check that the following subspace
\begin{equation}
\label{si}
(\hat N_i-s_i)\phi=0
\end{equation}
is $\Omega-$invariant so that one can define $\Omega-$cohomology in the subspace~\eqref{si}.

\begin{lemma}
\label{lemma:upp-triang}
Let weights $s_i$ be such that $s_1\geq s_2 \geq\ldots\geq s_{n-1}$ (i.e. respective $sl(n-1)$-weight are nonnegative)
then $\Omega$-cohomology vanishes in nonzero degree.
\end{lemma}
Note that the cohomology at vanishing degree is clearly a subspace of vectors satisfying the
highest-weight condition, i.e. vectors annihilated by all $N_i^j$ with $i<j$.
\begin{proof}
The statement can be proved by induction. The first nontrivial case is $n=2$ where the statement immediately follows
from the structure of irreducible $sl(2)$-modules. Suppose it is true for $n=k$. If $\Omega_k$ is the respective BRST operator then $\Omega_{k+1}$ takes the form
\begin{equation}
\Omega_{k+1}=\Omega_k+c^l N_{l}^{k+1}-\gamma^j_i(c^i\dl{c^j})\,,
\end{equation}
where there is no summation over $k$ and summation over $i,j,l$ runs from $1$ to $k$. We also introduced notations $c^i$ for $\gamma^i_{k+1}$. It can be rewritten as $\Omega_{k+1}=\hat\Omega_{k}+c^l N_{l}^{k+1}$,
where $\hat\Omega_k$ is obtained from $\Omega_k$ by replacing $N_i^j$ with $\cN_i^j=N_i^j-c^j\dl{c^i}$.
New generators form the same algebra so that $\hat\Omega_k$ is also a BRST operator of the same upper-triangular subalgebra but with coefficients in a different finite-dimensional representation. In particular, $\hat\Omega_k$ is nilpotent. Moreover, the induction assumption is satisfied for $\hat\Omega_k$ acting in this representation. Note that the weight conditions
take the form $(\cN_i-c^i\dl{c^i}-s_i)\Psi=0$ in this case, where, again, no summation over $i$ is assumed.

$\Omega_{k+1}$-cohomology can be computed as follows. Taking as a degree minus homogeneity in $\gamma$
one finds that $\hat\Omega_{k}$ is the lowest (degree $-1$) term in $\Omega_{k+1}$. The cohomological problem can be reduced to its cohomology. By the indiction assumption cohomology of $\hat\Omega_k$ is concentrated
in degree zero (are given by $\gamma$-independent elements annihilated by $\cN_i^j$ with $i<j$). The cohomology problem reduces then to the cohomology of $c^l N_{l}^{k+1}$ in this subspace. But this problem is identical to that considered in~\cite{Alkalaev:2008gi}. Using this result and taking into account the weight condition $s_{k+1}\leq s_i$ one concludes that the cohomology is given by $c^i$-independent elements
annihilated by $N_{l}^{k+1}$ so that the statement remains true at the next step of the induction.
\end{proof}

The above statatement underlies the equivalence of the parent formulation based
on $\bar\Omega^{\rm tot}$ and the formulation based on BRST operator
$Q_p$ and the constraints~\eqref{mainconst-fiber}-\eqref{newS-fiber}.
Indeed, the term $c_{IJ} T^{IJ}$ in $\bar\Omega^{\rm tot}$  implements tracelessness conditions. Reducing
to its cohomology simply amounts to eliminating ghosts $c_{IJ}$ and assuming all
elements totally traceless (see~\cite{\AGT} for details and proofs). Using then
a suitable degree one can assume that the term $\nu_i\bsd^i+\mu
(h+\text{ghosts})$ has the lowest degree. Its cohomology can be identified with
$\mu,\nu,\uu_i,z$-independent elements (see Section~\bref{sec:lorentz} and
Lemma~\bref{prop:isom} for notation and further details). In terms of this
identification  constraints $N_i{}^j$ act as $n_i{}^j=a^a_i\dl{a^a_j}$, while
$Q_p$ acts as
$q_p=s^\dagger_\alpha\dl{b_\alpha}=a^a_\alpha\dl{y^a}\dl{b_\alpha}$. The above
steps are identical to those in~\cite{\AG} to which we refer for further
details.

As a next step one takes as a degree $\deg\gamma^i{}_j=-1$ so that the term implementing $n_i^j$
has the lowest degree and we reduce the formulation to its cohomology.  It follows from
Lemma~\bref{lemma:upp-triang} that the cohomology can be taken $\gamma$-independent. In this way one reduces the formulation to that based on $q_p$. At the same time, starting with the formulation based on $Q_p$ and following~\cite{\AG} (or equivalently, specializing the reduction described in~\bref{sec:lorentz}) one arrives at the same formulation by explicitly solving $\bsd^i,h$ constraints.

\section{AdS Weyl module: technical details}
\label{sec:AppendixA}

In this Appendix we collect various technical details needed for the discussion of AdS Weyl module in
Section~\bref{sec:weylmodule}.

\begin{lemma}
\label{prop:isom}
 The space of all totally traceless elements $\Psi = \Psi(Y,A)$ satisfying
 \begin{equation}
\label{bsdh}
(\bsd^{p})^t \Psi = 0\;,
\qquad
\bsd^{\hat \alpha} \Psi = 0\;,
\qquad
h\Psi=0\,,
\end{equation}
where $\hat\alpha =  p+1,..., n-1$,
is isomorphic to the space of totally traceless
elements $\Psi = \Psi(a,y,w,z)$ satisfying
\begin{equation}
\label{diffree}
\Big(\frac{\d }{\d \uu_p}\Big)^{t}\Psi = 0\;,
\qquad
\frac{\d }{\d \uu_{\hat \alpha}}\Psi = 0\;,
\qquad
\frac{\d }{\d z}\Psi = 0\;.
\end{equation}
\end{lemma}
\noindent
The proof is a straightforward generalization of that
from~\cite{\BGadS,Alkalaev:2009vm}. The only modification has to do with
taking traces into account. To generalize the recursive proof of~\cite{\BGadS,Alkalaev:2009vm}
to the present case one needs to show that the cohomology of the auxiliary BRST operator
$\delta=\mu\dl{z}+\nu_{\hat\alpha}\dl{\uu_{\hat\alpha}}+\nu_p (\dl{\uu_p})^t+C_{IJ}T^{IJ}$
is trivial at nonvanishing degree in auxiliary ghost variables $\mu,\nu,C$. To see this one
first reduces to cohomology of $\mu\dl{z}+\nu_{\hat\alpha}\dl{\uu_{\hat \alpha}}+\nu_p (\dl{\uu_p})^t$ and hence eliminates ghosts
$\mu,\nu$. Using then a degree such that $\deg{\uu_i}=-1$ the lowest degree term of the reduced differential is
simply $C_{IJ}T_0^{IJ}$ where $T_0^{IJ}$ is obtained from $T^{IJ}$ by omitting terms involving $\dl{z}$ and $\dl{\uu_i}$.
Finally, $T_0^{IJ}$ are usual trace operators in $d$ dimensions and hence cohomology of $C_{IJ}T_0^{IJ}$
is concentrated at zeroth ghost degree for ghosts $C_{IJ}$~\cite{\AGT}. As ghost variables $\mu,\nu$
have been already eliminated at the previous step one concludes that cohomology is trivial at nonvanishing
degree in auxiliary ghost variables.

The next fact we need is the explicit solution to irreducibility conditions
 \eqref{reduYoung}.  Namely, the  space of solutions to~\eqref{reduYoung}
can be isomorphically mapped to the subspace singled out by
\begin{equation}
\label{reduyoungnew}
\big(n_i{}^j + \delta_i^\alpha\delta^j_\beta \uu_{\alpha}\dl{\uu_{\beta}}+\delta_i^\alpha\delta^j_\beta B_\alpha{}^\beta\big) \tilde \psi = 0\,,
\qquad
i<j\;.
\end{equation}
This can be shown by analyzing the recurrent equations
obtained by substituting level-$k$ decomposition
\eqref{decW} into \eqref{reduYoung}. Using decomposition \eqref{decW} and  separating the term
$-\uu_\alpha \bar{s}^{\,\hat \beta}$ by prescribing $\uu_\alpha$ to carry degree $1$
one recursively shows that a space of solutions to modified Young conditions
\eqref{reduYoung} can be mapped to the subspace  of
elements satisfying \eqref{reduyoungnew}.

In its turn the subspace \eqref{reduyoungnew} can be isomorphically
mapped to the following subspace:
\begin{equation}
\label{reduyoungnewNEW}
\big(n_i{}^j+\delta_i^\alpha\delta^j_\beta B_\alpha{}^\beta\big) \hat \psi = 0\,,
\qquad
i<j\;.
\end{equation}
To see this one again substitutes decomposition
\eqref{decW} into \eqref{reduyoungnew}. Solution to the resulting
inhomogeneous linear equations are parameterized
by elements satisfying  \eqref{reduyoungnewNEW}.
It is important to stress that the $q_p$
represented in terms of parameterization~\eqref{reduyoungnewNEW}
remains intact because it commutes both with $\uu_\alpha$-variables
and BRST extended Young conditions \eqref{reduyoungnewNEW}.

\paragraph{Proof of Proposition \bref{qp}.}
\label{proof42} Using the above isomorphisms reduces the problem to calculating
$q_p$-cohomology in the subspace \eqref{reduyoungnewNEW}.
The $q_p$-cohomology problem in the subspace \eqref{reduyoungnewNEW}
is identical to that considered in~~\cite{Alkalaev:2009vm}.
Applying then lemmas \textbf{5.3} and \textbf{5.4} from~\cite{Alkalaev:2009vm}
gives the statement.

\paragraph{Proof of Proposition \bref{prop:WEYL}.}
\label{proof43}

The zero-ghost-number cohomology of the total BRST operator $Q_p$ is defined by the
following chain of equivalence relations read off from
\eqref{decW} and \eqref{lorq}:
\begin{equation}
\label{quotients}
\ba{l}
\dps
\psi^{\alpha_1... \alpha_k}_k \;\sim\; \psi^{\alpha_1... \alpha_k}_k
+ \smalls^\dagger_\gamma \chi_k^{\alpha_1... \alpha_k|\gamma} - \hat h \chi_{k-1}^{\alpha_1... \alpha_{k-1}|\alpha_k}\;.
\ea
\end{equation}
To fix representatives one proceeds as follows. First of all one finds representatives
of the above relations without the term containing $\hat h$. These are simply representatives
of $q_p$-cohomology at zeroth ghost number described by Proposition \bref{qp}.
Taking $\hat h$ into account  amounts to subtracting particular components
from $H^0(q_p)$. To clarify which  components should be cancelled out
one analyzes the following residual equivalence condition:
\begin{equation}
\smalls^\dagger_\gamma \chi_{k}^{\alpha_1 ... \alpha_{k}|\gamma}-\hat h \chi_{k-1}^{\alpha_1 ... \alpha_{k-1}|\alpha_k}=0\;.
\end{equation}
One then observes that admissible
$\chi_{k}^{\alpha_1 ... \alpha_{k}|\gamma}$ are all described by Young diagrams with
$\# y^a  \geq \# a_1^a  $, where $\#$ denote the homogeneity degree in the respective variable.
More precisely, using a technique elaborated in \cite{Alkalaev:2009vm}  one shows that
for the $k$-th level the homogeneity in $y^a$ variables
is $s_1-k \leq \# y^a \leq s_1-1$. The remaining weight and Young conditions
imposed on $\chi_{k}^{\alpha_1 ... \alpha_{k}|\gamma}$ are such that both sides of
the equivalence relations \eqref{quotients} satisfy the same algebraic
constraints.

\paragraph{An example: $n=2\,$, $\,p=1\,$, $\,t\geq 1$.}
\label{sec:example}

In what follows we explicitly demonstrate the  $Q_p$-cohomology calculation
for the simplest case of totally-symmetric
partially-massless fields of spin $s$ and depth $t$~\cite{Deser:2001pe,Deser:2001us,Zinoviev:2001dt,Skvortsov:2006at}.

Decomposition~\eqref{decW} takes the following form
%
\begin{equation}
\label{psis}
\psi(a,y,w|b) = \sum_{k=0}^{t-1}\psi_k(a,y|b)\uu^k\;.
\end{equation}
BRST operator $\tilde Q$ is given by
\begin{equation}
\tilde Q_p=s^\dagger \dl{b} - \uu \hat h \dl{b} \equiv q_p - \uu \hat h \dl{b}\;,
\end{equation}
where $\hat h  = n_y -B -s+t +1$.
It acts in the subspace of \eqref{psis} singled out by the weight constraints
\begin{equation}
\ba{c}
\dps
\big(n_a + n_\uu+ n_b-s\big)\psi(a,y,\uu|b) = 0\;.
\ea
\end{equation}

Let us analyze first the cohomology in the minimal ghost number $-1$. For $\psi  = b\psi^1$
we get
\begin{equation}
s^\dagger \psi^1_k - \hat h \psi_{k-1}^1 =
0\;, \qquad k=0,...,t-1\;.
\end{equation}
It follows that $\psi^1_k$
consists of two parts: the kernel of $s^\dagger$
and the particular solution determined by $\psi_{k-1}^1$. The exact formula reads
as
\begin{equation}
\psi^1_{k} = \tilde{\psi}_{k}^1 + \bar{s} \frac{\hat h\, \psi_{k-1}^1}{n_a-n_y}\;,
\qquad \text{where} \qquad s^\dagger \tilde{\psi}_{k}^1 = 0\;.
\end{equation}
It follows that for some elements  of level $k-1$ the denominator
may vanish. This implies that these elements are set to zero.

Analyzing  the above system of equations recursively results in  a set of Lorentz
components defined by constraints
\begin{equation}
s^\dagger \tilde{\psi}^1_k = 0\;,
\qquad
(n_a-s+k+1) \tilde{\psi}^1_k = 0\;,
\qquad
(n_y-l) \tilde{\psi}^1_k = 0\;,
\end{equation}
for
\begin{equation}
k = 0, ... , t-1\;, \qquad l = 0,...,s- t\;.
\end{equation}
In other words, they are described by diagrams with two rows, the first one is of
length $s-k-1$ and the length of the second row is not exceeding $s-t$.
In manifestly $o(d-1,2)$ terms these are describes by a single two-row
diagram with the lengths of rows $s-1$ and $s-t$ (see Section \bref{sec:gauge module}).

Representing the gauge parameter as
$\dps\chi^1 =
b\sum_{k=0}^{t-1} \chi^1_k(a,y) w^k $ we  cast
the cocycle condition into the following form
\begin{equation}
\psi^0_k \sim \psi^0_k + s^\dagger \chi^1_k - \hat h \chi^1_{k-1}\;,
\qquad
k = 0,..., t-1\;.
\end{equation}
The term $-\hat h \chi^1_{k-1}$ defines
Stueckelberg-like transformation with parameter $\chi^1_{k-1}$ satisfying
the gauge fixing condition $s^\dagger \chi^1_{k-1} - \hat h \chi^1_{k-2}=0$.

To identify representatives of the above equivalence relations we analyze them
recursively starting from the level $k=0$.
The end result is the following collection of Lorentz tensors:
\begin{equation}
\psi_k^{a(s+l-k),\, b(s-k)}\;,\quad \qquad k=0,...,t-1\;,\qquad l\geq k\;,
\end{equation}
where (using the notation of Ref.~\cite{Vasiliev:1986td}) a set of $s$ symmetrized indices $a$ is denoted by $a(s)$ while different groups of symmetrized indices are separated by a comma.
This describes AdS Weyl module for spin $s$ and depth $t$ gauge field. Note that it can be also described in manifestly $o(d-1,2)$ covariant notation, see~\cite{Alkalaev:2009vm,Skvortsov:2009nv}.

\providecommand{\href}[2]{#2}\begingroup\raggedright
\addtolength{\baselineskip}{-5pt}
\addtolength{\parskip}{-2pt}
\providecommand{\href}[2]{#2}\begingroup\raggedright\endgroup


\end{document}